\newif\ifmulticol	\multicoltrue
\newif\ifshowgit	\showgittrue		
\newif\ifgitlocal	\gitlocaltrue		
\newif\ifbiblatex	\biblatexfalse		
\newif\ifbibnum		\bibnumtrue 		
\newif\ifbibsort	\bibsortfalse		
\newif\iflineno		\linenofalse
\newif\iftoc		\tocfalse

\newif\iflucida		\lucidafalse
\newif\ifcm			\cmfalse
\newif\iflibertine	\libertinefalse		
\newif\ifcharter	\chartertrue

\newcommand*{\secnumdepth}{0}			
\newcommand*{\tocdepth}{2}				
\newcommand*{\reftitle}{References}		
\newcommand*{\mytitleskip}{-0.2in}		

\newcommand*{\mydocfontsize}{\ifcharter11pt\else\iflibertine11pt\else10pt\fi\fi}
\newcommand*{\setcol}{\ifmulticol twocolumn\else onecolumn\fi}

\documentclass[\mydocfontsize,\setcol]{article}

\newcommand*{\pdfauthor}{Steven A.~Frank}
\newcommand*{\pdftitle}{A biological circuit to anticipate trend}

\usepackage{afterpage}

\input trend.sty

\newcommand*{\Ga}{\alpha}
\newcommand*{\Gb}{\beta}
\newcommand*{\Gd}{\delta}
\newcommand*{\GD}{\Delta}

\newcommand*{\Gg}{\gamma}

\newcommand*{\Gl}{\lambda}

\newcommand*{\Gs}{\sigma}
\newcommand*{\Gt}{\tau}

\usepackage{bm}

\DeclarePairedDelimiter\abs{\lvert}{\rvert}
\DeclarePairedDelimiter\norm{\lVert}{\rVert}
\DeclarePairedDelimiter\angb{\langle}{\rangle}
\DeclarePairedDelimiter\lrb{\lbrack}{\rbrack}
\DeclarePairedDelimiter\lr{\lparen}{\rparen}
\DeclarePairedDelimiter\lrbr{\lbrace}{\rbrace}

\makeatletter
\let\oldabs\abs \def\abs{\@ifstar{\oldabs}{\oldabs*}}
\let\oldnorm\norm \def\norm{\@ifstar{\oldnorm}{\oldnorm*}}
\let\oldangb\angb \def\angb{\@ifstar{\oldangb}{\oldangb*}}
\let\oldlrb\lrb \def\lrb{\@ifstar{\oldlrb}{\oldlrb*}}
\let\oldlr\lr \def\lr{\@ifstar{\oldlr}{\oldlr*}}
\let\oldlrbr\lrbr \def\lrbr{\@ifstar{\oldlrbr}{\oldlrbr*}}
\makeatother

\newcommand*{\dd}{\textrm{d}}

\newcommand*{\Eq}[1]{eqn~\ref{eq:#1}}

\newcommand*{\boldrule}{\hrule height 1.2pt}
\newcommand*{\noterule}{\medskip\boldrule\medskip}	

\hypersetup{draft}

\newcommand*{\mytitle}{\pdftitle}

\newcommand*{\myauthor}{Steven A.\ Frank}
\newcommand*{\myaddress}{Department of Ecology and Evolutionary Biology, University of California, Irvine, CA 92697--2525, USA}
\newcommand*{\mycontact}{\href{https://stevefrank.org}{https://stevefrank.org}}

\newcommand*{\mykeywords}{Evolutionary theory, plasticity, systems biology, financial models, stochastic time series}

\setabstract{-0.07}{\iftoc-2.0\else1.22\fi}{%
Organisms gain by anticipating future changes in the environment. Those environmental changes often follow stochastic trends. The greater the slope of the trend, the more likely the trend's momentum carries the future trend in the same direction. This article presents a simple biological circuit that measures the momentum, providing a prediction about future trend. The circuit calculates the momentum by the difference between a short-term and a long-term exponential moving average. The time lengths of the two moving averages can be adjusted by changing the decay rates of state variables. Different time lengths for those averages trade off between errors caused by noise and errors caused by lags in predicting a change in the direction of the trend. Prior studies have emphasized circuits that make similar calculations about trends. However, those prior studies embedded their analyses in the details of particular applications, obscuring the simple generality and wide applicability of the approach. The model here contributes to the topic by clarifying the great simplicity and generality of anticipation for stochastic trends. This article also notes that, in financial analysis, the difference between moving averages is widely used to predict future trends in asset prices. The financial measure is called the moving average convergence-divergence (MACD) indicator. Connecting the biological problem to financial analysis opens the way for future studies in biology to exploit the variety of highly developed trend models in finance. 
}

\begin{document}

\mymaketitle

\iftoc\mytoc{-24pt}{\newpage}\fi

\section{Introduction}

Predicting future environmental change provides many benefits. A microbe gains by anticipating the availability of sugars. A plant gains by forecasting the flow of nutrients. Plasticity benefits from a head start on altering physiology or form. Competitive players profit by preparing for the next step in a contest \autocite{pigliucci01phenotypic,dewitt04phenotypic,kussell05phenotypic,mitchell09adaptive,reed10phenotypic,shemesh10anticipating,goo12bacterial,siegal15shifting,venturelli15population,21phenotypic}.

Organisms often anticipate regular patterns of change, such as circadian rhythms or seasonal cycles. However, many regularities arise as stochastic trends. To take advantage of such trends, a biological circuit must measure past directionality and use that measurement to predict the direction of future change.

In the literature, studies of \textit{E.\ coli} chemotaxis develop the most compelling models for the anticipation of trend \autocite{tu08modeling,shimizu10amodular,alon19an-introduction}. Cells measure changes in chemical concentrations to predict whether future changes will be increasing or decreasing. Tjalma et al.'s \autocite{tjalma23trade-offs} excellent recent article synthesizes past literature and develops new models.

This article introduces a simple model for anticipating trend. Roughly speaking, the model estimates the momentum of the current trend by the difference between a shorter-term moving average and a longer-term moving average. That estimate of momentum predicts the future direction of change because the future trend often continues in the direction of the current momentum.

Most existing models, such as those for \textit{E.\ coli} chemotaxis, also base predictions on estimates of current or past trend. However, those previous models typically add specific aspects of a particular application or additional complexities, such as noise filtering. Those additions are interesting but also obscure the simplicity and generality of the underlying way in which estimates of momentum anticipate future trend.

The model here strips nonessential features to emphasize the fundamental structure of anticipation for simple stochastic trends. That abstraction provides the basis for future applications across a broader range of biological problems.

Empirically, the model predicts the primary mechanism that organisms use to anticipate environmental trends and the pattern of anticipatory response to environmental changes. Theoretically, the model provides the foundation for developing further predictions about tradeoffs between rapid adjustment of anticipated environmental changes and the susceptibility to perturbation of the circuit that predicts trends \autocite{tjalma23trade-offs}.

The model's simplicity also reveals a close connection to the moving average convergence-divergence (MACD) indicator, the most widely used measure of momentum and trend to analyze asset prices in financial time series \autocite{murphy99technical,schwager99getting,pring14technical}. That connection between biological models and financial analysis encourages application of the highly developed models of information and anticipation in finance to biological problems.

\section{The challenge}

Let $u_t$ be a randomly varying input signal. We wish to predict the direction of change at a future time, $t+1$, relative to the current time, $t$, which means predicting the sign of $\GD u_t = u_{t+1}-u_t$. If $u$ changes in a purely random way, as a random walk with no directionality, then expected prediction success above $1/2$ is not possible. For prediction to be possible, we must assume some pattern to the fluctuations in $u_t$.

An exponential moving average of purely random inputs is perhaps the most generic type of trend. Start with a random walk, $\hat{u}_t$, sampled at discrete times, $t=0,1,\dots$. Then replace each value at time $t$ by its exponential moving average, $u_t = (1-\Gd) \hat{u}_t + \Gd u_{t-1}$. The value of $0\le\Gd\le1$ sets the memory scale, with larger values averaging over longer time periods. Here, we linearly interpolate between the discretely sampled points to obtain values of $u_t$ continuously with time.

The future change in $u_t$ is positive when
\begin{equation}\label{eq:ema}
  \GD u_t = (1-\Gd)\GD\hat{u}_t + \Gd\GD u_{t-1} > 0.
\end{equation}
Because $\hat{u}_t$ is a random walk, its change is equally likely to be positive or negative. Thus, the best prediction for the sign of $\GD u_t$ is the sign of $\GD u_{t-1}$. In other words, the most recently observed direction of change provides the best prediction for the next direction of change.

Roughly speaking, we may think of the currently observed change, $\GD u_{t-1}$, as the trend momentum. Positive momentum means that the trend is likely to continue up. Negative momentum means that the trend is likely to continue down.

The greater the momentum, the more likely the trend will continue in the same direction. For example, the larger $\GD u_{t-1}$, the more negative the underlying random walk change, $\GD\hat{u}_t$, must be to reverse the trend. Increasingly extreme moves in the underlying random walk are increasingly uncommon.

Thus, an ideal internal model uses the currently observed trend direction to predict the next direction of change. And it uses the trend momentum to estimate the confidence in the directional prediction.

\section{Exponential moving average}

I claimed that an exponential moving average provides a common type of trend. I made that claim because any process that balances a fluctuating input against a steady decay describes an exponential moving average. For example, the differential equation
\begin{equation*}
  \dot{z} = u - \Gl z
\end{equation*}
determines a level of $z$ that balances production or input, $u$, against steady decay at rate $\Gl$. With $u_0=0$, the value of $z$ at time $t$ is
\begin{equation*}
  z(t) = \int_0^t e^{-\Gl(t-\Gt)}u(\Gt)\dd\Gt,
\end{equation*}
which is proportional to the continuous time exponential moving average of $z$ at that time. The process in \Eq{ema} is a discrete time version of the balance between input and decay. 

\section{The circuit}

To predict the trend direction, the following circuit calculates the difference between two moving averages, each average taken over a different time scale
\begin{align}\label{eq:cellTrend}
\begin{split}
  \dot{x}&=\Ga u - \Gb x\\
  \dot{y}&=\Gg\lr{\Ga u - \Gb y}.
\end{split}
\end{align}
Overdots denote derivatives with respect to time, $t$, and $x,y,u$ are functions of time.
For constant input, $u$, the molecular of abundances of $x$ and $y$ have the same equilibrium values, $\Ga/\Gb$. As the input, $u$, changes, $y$ changes more slowly than $x$ when $\Gg < 1$. Thus, $x-y$ tends to be positive when $u$ is increasing and negative when $u$ is decreasing. The values of $x$ and $y$ are approximately proportional to shorter and longer exponential moving averages of the input, $u$.

\section{An example}

We can optimize the parameters relative to a particular goal. Figure 1 shows an example of an optimized circuit that predicts the future direction of change in $u_t$. Define maximum potential accuracy as the frequency at which the prior sign of change in $u_t$ predicts the next sign of change, a continuation of trend.

In Fig.~1, the maximum potential accuracy is $0.8$. Stochasticity in the input sequence means that sometimes the sign of the prior difference does not match the direction of the next difference. For this example, the optimized circuit accuracy is close the maximum potential accuracy, with a median deviation from the maximum of $0.001$. 

Figure 1b shows the classic MACD pattern for trend prediction from financial analysis. The slower signal in gold, which is the variable $y$, lags the faster signal in blue, which is the variable, $x$. A predicted trend reversal arises when the fast blue signal crosses the slow gold signal, that is, when the sign of $x-y$ changes. Figure 1c traces a function of the $x-y$ difference.

A convergence between the trends in panel b and c predicts a continuation of the current trend. A divergence between the trends in the two panels foreshadows a potential upcoming change in direction for the trend. 

For example, starting at time point 98, the $x-y$ difference begins to shrink, causing an upturn in the trend in panel c. That uptrend signals a slowing of the downward momentum. At first the actual trend in panel b continues downward. Thus, the trends in the two panels are diverging. Then, at time point 104, the actual trend in b turns up, a convergence with the trend in c, signaling the potential start of an upswing, which in fact does occur.

Typically, a change in momentum (panel c) precedes a change in trend (panel b), as happens around time point 104. The reason a slowing of momentum typically precedes a change in trend can be seen in \Eq{ema}, which shows how changes in the direction of an exponentially smoothed input are often dominated by the momentum term. However, when the trend changes abruptly, momentum does not precede trend, as happends around time point 92.

\section{Accuracy versus robustness}

High accuracy requires that the $x-y$ predictor in Fig.~1c switches sign immediately when the input turns down. The input appears as the red line in Fig.~1a. To achieve high accuracy in forecasting a change in trend direction, in Fig.~1b, the slower moving average, $y$, in gold, must remain close to the faster moving average, $x$, in blue.

The small difference between the moving averages, $x$ and $y$, means that false signals can arise from small perturbations. For example, at time point 102 in Fig.~1b, the moving averages touch, predicting an imminent reversal that does not subsequently occur. In addition, noise in the state variables $x$ and $y$ can trigger false signals.

Figure 2 shows the circuit optimized for wider differences between the fast and slow moving averages, $x$ and $y$. In panel b, The wider differences provide a stronger signal during a trend but lag in giving a signal when the trend does change direction. This system gives fewer false signals and is robust to small perturbations. However, because it lags when the trend does change, it has a lower accuracy when using the sign of the current difference, $x-y$, to predict the direction of change in the input in the next time period. The median deviation from the maximum accuracy is $0.12$, compared with a deviation of $0.001$ in the circuit illustrated in Fig.~1. 

In general, systems can be tuned to balance various tradeoffs, such as accuracy versus robustness \autocite{murphy99technical,schwager99getting,pring14technical,tjalma23trade-offs}.

\begin{figure*}[t]
\centering
\includegraphics[width=0.7\hsize]{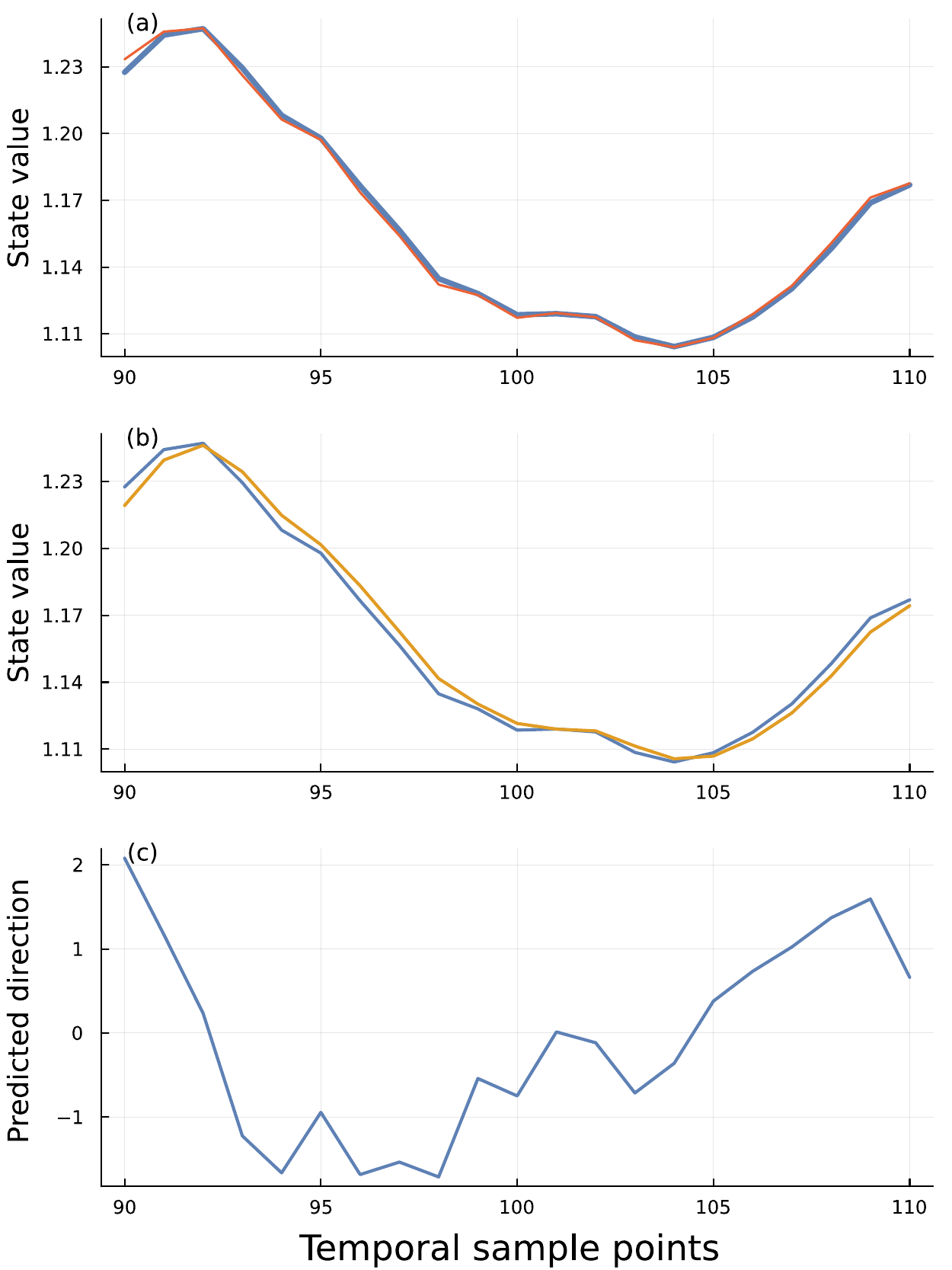}
\vskip7pt
\caption{Prediction for the direction of change in a sequence of observations. Dynamics given by \Eq{cellTrend}. In this example, the parameters $\Ga,\Gb,\Gg$ were optimized for accuracy of prediction, yielding $0.1663,0.08314,0.2782$, respectively. The plots show a subset of time points to magnify patterns and make them easier to see. (a) The faster of the two moving averages, $x$, in blue, tracks the input sequence, $u_t$, in red. (b) The fast and slow moving averages, $x$ and $y$, in blue and gold, respectively. (c) The difference between the moving averages, plotted as $1000[\Gs(x-y)-0.5]$, in which $\Gs(z)=e^z\large/\large(1+e^z\large)$ is the sigmoid function. The sign predicts the direction of change in the next time step. The magnitude reflects the momentum, a measure of the relative confidence in the prediction for the future direction of change. To calculate the input sequence in \Eq{ema}, the random walk follows a Wiener process with mean $0$ and standard deviation $0.2$, and the exponential moving average memory parameter is $\Gd=0.2$. For each realized sequence of the random walk, the values are normalized to $[0.25,0.75]$ by affine transformation, yielding $\hat{u}$. The timescale and abundances in the plots have arbitrary units. The freely available computer code describes the scaling of time, the parameters, the optimization, and the production of graphics \autocite{steven_a_frank_2024_10967108}.
}
\label{fig:cellTrend}
\end{figure*}

\begin{figure*}[t]
\centering
\includegraphics[width=0.7\hsize]{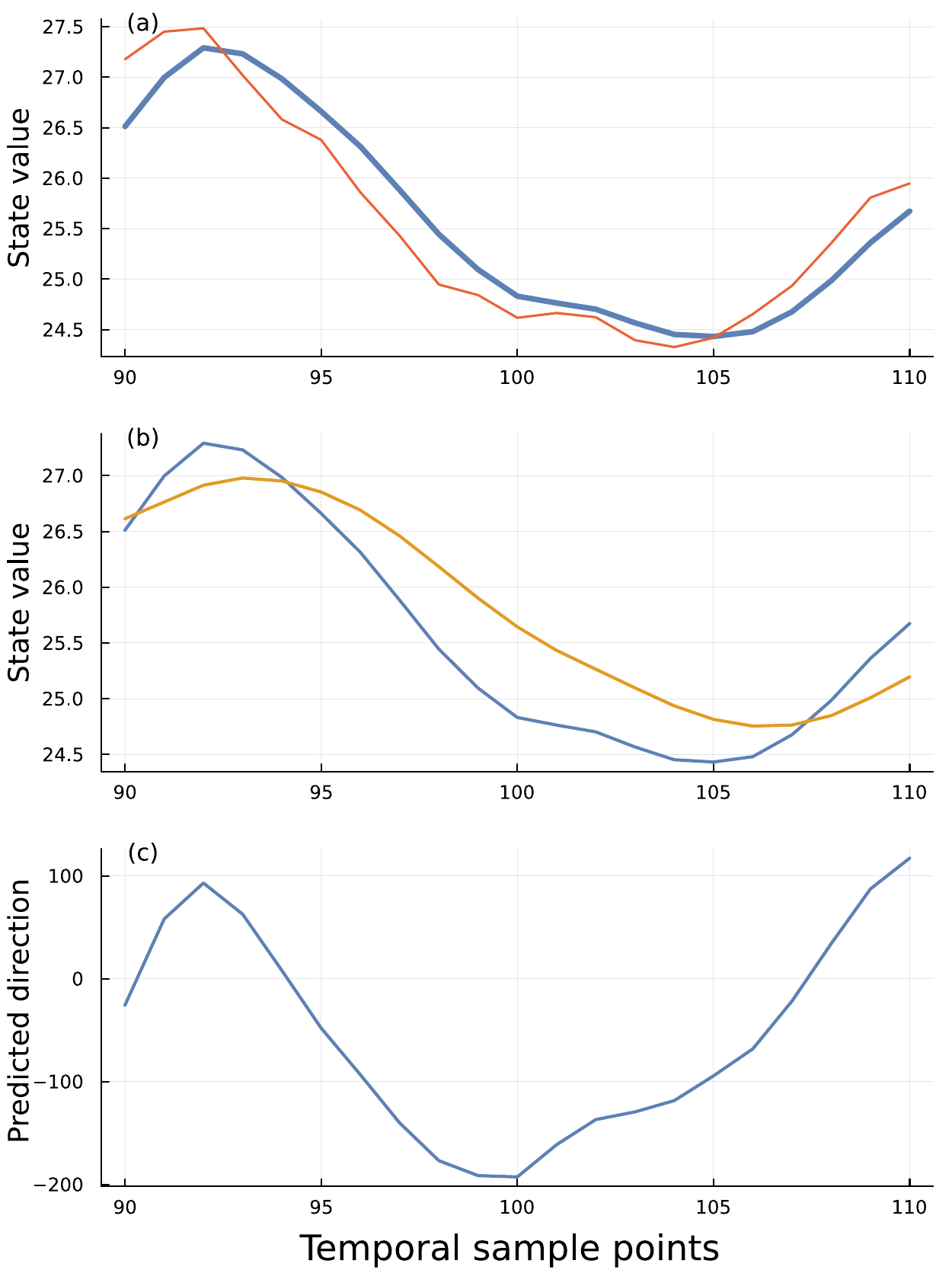}
\vskip7pt
\caption{Tradeoff between accuracy and robustness. Same plots as Fig.~1 but with slower moving averages. Optimized parameters $\Ga,\Gb,\Gg$ are $0.4179,0.009483,0.2587$.
}
\label{fig:cellTrend2}
\end{figure*}

\section{Conclusion}

Prior models analyzed how biochemical circuits predict future environmental changes \autocite{tu08modeling,shimizu10amodular,alon19an-introduction,mitchell09adaptive,becker15optimal,tjalma23trade-offs}. Those models use recent differences in input to predict future changes because that is the essential nature of the problem.

Although the prior models described biological circuits that predict environmental changes, none of those prior analyses presented a circuit and an explanation as simple as those given here. In essence, the difference between a shorter, more immediate moving average, $x$, and a longer, slower moving average, $y$, provides the basis to forecast trends.

In the technical analysis of financial prices, the MACD indicator for moving average convergence-divergence is calculated as the difference between longer and shorter moving averages. Organisms may use a similar calculation to anticipate environmental trends.

\section*{Author contributions}

SAF did everything.

\section*{Acknowledgments}

\noindent The Donald Bren Foundation, US Department of Defense grant W911NF2010227, and US National Science Foundation grant DEB-2325755 support my research.

\section*{Supplemental information}

Software code and output for all analyses and figures available at GitHub \autocite{steven_a_frank_2024_10967108}.


\mybiblio	


\end{document}